\documentclass[preprint,12pt]{elsarticle}

\usepackage{amssymb}
\usepackage{amsmath}

\usepackage{graphicx}
\DeclareGraphicsExtensions{.pdf,.png,.jpg,.eps}

\newtheorem{theorem}{Theorem}

\newtheorem{definition}{Definition}

\journal{Systems $\&$ Control Letters}

\begin{document}

\begin{frontmatter}



\title{Divergence Conditions for Investigation and Control of Nonautonomous Dynamical Systems$^\star$}


\author{Igor Furtat$^{a,b,*}$} 

\address{$^{a}$Institute for Problems of Mechanical Engineering Russian Academy of Sciences, 61 Bolshoy ave V.O., St.-Petersburg, 199178, Russia}
\address{$^{b}$ITMO University, 49 Kronverkskiy ave, Saint Petersburg, 197101, Russia}

\begin{abstract}
The paper describes a novel method for studying the stability of nonautonomous dynamical systems.
This method based on the flow and divergence of the vector field with coupling to the method of Lyapunov functions.
The necessary and sufficient stability conditions are formulated.
It is shown that the necessary stability conditions are related to the integral and differential forms of continuity equations with the sources (the flux is directed inward) located in the equilibrium points of the dynamical system.
The sufficient stability conditions are applied to design the state feedback control laws.
The proposed control law is found as a solution of partial differential inequality, whereas the control law based on Lyapunov technique is found from the solution of algebraic inequality.
The examples illustrate the effectiveness of the proposed method compared with some existing ones.
\end{abstract}



\begin{keyword}
Nonautonomous dynamical system, stability, flow of vector field, divergence, control.
\end{keyword}

\end{frontmatter}



\section{Introduction}

The method of Lyapunov functions is a powerful tool for studying the stability of solutions of differential equations without solving them. Depending on the problem being solved, Lyapunov function is also interpreted as a potential function \cite{Yuan14}, an energy function \cite{Bikdash00} or a storage function \cite{Willems72}. The main restriction of the method of Lyapunov functions is to find these functions.

Methods for stability study of dynamical systems based on the divergence of a vector field are alternative to the method of Lyapunov functions. The first fundamental results based on divergent stability conditions were proposed in \cite{Zaremba54,Fronteau65,Brauchli68}. 
The last important results for investigation of system stability were proposed by A. Rantzer, A.A. Shestakov,  A.N. Stepanov and V.P. Zhukov.
In \cite{Zhukov78} the instability problem of nonlinear systems using the divergence of a vector field is considered. 
In \cite{Shestakov78,Zhukov79} a necessary condition for stability of nonlinear systems in the form of non-positivity of the vector field divergence is proposed. 
First, an auxiliary scalar function is introduced in \cite{Shestakov78,Zhukov90} to study the instability of nonlinear systems. 
However, the similar scalar function is considered in \cite{Krasnoselski63} for stability and instability study of dynamical systems, but using the method of Lyapunov functions.
In \cite{Shestakov78,Zhukov99} stability conditions for second-order systems are obtained. 
Then in \cite{Rantzer00,Rantzer01} the convergence of almost all solutions of arbitrary order nonlinear dynamical systems is considered. 
As in \cite{Shestakov78,Zhukov90,Zhukov99} the auxiliary scalar function (density function) is used for the stability study of dynamical models. Additionally, in \cite{Rantzer00,Rantzer01} the synthesis of the control law based on divergence conditions is proposed.
The auxiliary functions in \cite{Shestakov78,Zhukov99,Rantzer00,Rantzer01} are similar except their properties at the equilibrium point. 
Currently, method from \cite{Rantzer00,Rantzer01} has been extended to various systems, see i.e. \cite{Monzon03,Loizou08,Castaneda15,Karabacak18}.

However, in \cite{Zaremba54,Shestakov78,Zhukov99} the necessary condition is sufficiently rough and it is obtained only for autonomous systems. 
The sufficient condition stability is proposed only for second-order autonomous systems in \cite{Zhukov99}. 
Corollary 1 in \cite{Rantzer01} guarantees the convergence of almost all solutions, but not all solutions, for nonautonomous systems. 
Proposition 2 in \cite{Rantzer01} allows to study the asymptotic stability for autonomous systems, but proposition conditions have sufficient restriction. 
In the present paper new necessary and sufficient conditions will be obtained that will eliminate the above disadvantages and expand the class of investigated systems.

In this paper a new method for the stability study of nonautonomous systems using the flow and divergence of the vector field is proposed.
The relation between the method of Lyapunov functions and the proposed method is established. 
The method for design the state feedback control law based on the new divergence conditions is proposed.
Numerical examples illustrate the applicability of the proposed method and the methods from \cite{Zaremba54,Shestakov78,Zhukov99,Rantzer00,Rantzer01}.

The paper is organized as follows. 
Section \ref{Sec2} contains new necessary and sufficient conditions, as well as, the numerical examples and comparisons with the methods from \cite{Zaremba54,Shestakov78,Zhukov99,Rantzer00,Rantzer01}. 
Section \ref{Sec3} describes methods for design the state feedback control law and numerical examples. 
Finally, Section \ref{Sec4} collects some conclusions.

\textit{Notations and definitions}. In the paper the following notation are used:  the superscript $\rm T$ stands for matrix transposition; $\mathbb R^{n}$ denotes the $n$ dimensional Euclidean space with vector norm $|\cdot|$; $\mathbb R^{n \times m}$ is the set of all $n \times m$ real matrices; $\nabla\{W(x,t)\}=\Big[\frac{\partial W}{\partial x_1}, ...,\frac{\partial W}{\partial x_n}\Big]^{\rm T} $ is the gradient of the scalar function $W(x,t)$, $\nabla \cdot\{h(x,t)\}=\frac{\partial h_1}{\partial x_1}+...+\frac{\partial h_n}{\partial x_n}$ is the divergence of the vector field $h(x,t)=[h_1(x,t),...,h(x,t)_n]^{\rm T}$, $|\cdot|$ is the Euclidean norm of the corresponding vector.

\begin{definition}
\label{Def1}
\cite{Khalil09}.
A continuous function $\alpha:[0,a) \to [0;\infty)$ is said to belong to class $\mathcal{K}$ if it is strictly increasing and $\alpha(0)=0$. 
\end{definition}

\begin{definition}
\label{Def2}
\cite{Khalil09}.
A continuous function $\beta:[0,a) \times [0;\infty) \to [0;\infty)$ is said to belong to class $\mathcal{KL}$ if, for each fixed s, the mapping $\beta(r,s)$ belongs to class $\mathcal{K}$ \textit{w.r.t.} $r$ and, for each fixed $r$, the mapping $\beta(r,s)$ is decreasing \textit{w.r.t.} $s$ and $\beta(r,s) \to 0$ as $s \to \infty$. 
\end{definition}

Additionally, in the paper we mean that the zero equilibrium point is stable if it is Lyapunov stable \cite{Khalil09}.

\section{Maun results} \label{Sec2}
Consider the nonautonomous system 
\begin{equation}
\label{eq1}
\begin{array}{l} 
\dot{x}=f(x,t),
\end{array}
\end{equation} 
where $x=[x_1, ..., x_n]^{\rm T}$ is the state vector, $f=[f_1,...,f_n]^{\rm T}: [0, \infty) \times D \to \mathbb R^{n}$ is  piecewise continuous in $t$ and continuously differentiable in $x$ on $[0, \infty) \times D$. 
The open set $D \subset \mathbb R^{n}$ contains the origin $x=0$ and $f(t,0)=0$ for any $t \geq 0$. 
Denote by $\bar{D}$ a boundary of the domain $D$.
Below, the structure of the set $D$ can be specified depending on the obtained result.

Let us formulate the necessary stability condition for system \eqref{eq1}.

\begin{theorem}
\label{Th1}
Let the Jacobian matrix $[\partial f / \partial x]$ be bounded on $D=\{x \in \mathbb R^n: \|x\|<r, r>0\}$ and uniformly in $t$, trajectories of system \eqref{eq1} satisfies $\|x(t)\| \leq \beta(\|x(t_0)\|,t-t_0)$ for any $x(t_0) \in D_0$ and $t \geq t_0 \geq 0$, where $\beta(\cdot,\cdot)$ is a class $\mathcal{KL}$ function, $D_0=\{x \in \mathbb R^n: \|x\|<r_0, r_0>0\}$ and $\beta(r_0,0)<r$.
Then there is a function $S(x,t): [0, \infty) \times D_0 \to \mathbb R$ 
such that $|\nabla\{S(x,t)\}| \neq 0$ for any $x \in D_0 \setminus \{0\}$, $t \geq 0$ 
and at least one of the following conditions holds:

\begin{enumerate}
\item[\textup{(1)}] the function 
$\frac{\partial S(x,t)}{\partial t}+\nabla \cdot\{|\nabla \{S(x,t)\} |f(x,t)\}$ 
is integrable in the domain 
$V=\{x \in D_0, t \geq 0: S(x,t) \leq C \}$ and 
$\int_{V} \Big[ \frac{\partial S(x,t)}{\partial t} + \nabla \cdot \{ |\nabla \{S(x,t)\}| f(x,t)\} \Big] dV <0$ for all $C>0$;

\item[\textup{(2)}] the function 
$\frac{\partial S^{-1}(x,t)}{\partial t}+\nabla \cdot \{|\nabla \{S^{-1}(x,t)\}|f(x,t)\}$ 
is integrable in the domain 
$V_{inv}=\{x \in D_0, t \geq 0: S^{-1}(x,t) \geq C \}$ and 
$\int_{V_{inv}} \Big[\frac{\partial S^{-1}(x,t)}{\partial t}+
\nabla \cdot\{|\nabla\{S^{-1}(x,t)\} |f(x,t)\}\Big] dV_{inv}>0$ for all $C>0$.
\end{enumerate}

\end{theorem}

\textbf{Proof 1}
According to \cite[Theorem 3.13]{Khalil09}, if Jacobian matrix $[\partial f / \partial x]$ is bounded on $D=\{x \in \mathbb R^n: \|x\|<r\}$ and uniformly in $t$, trajectories of  system \eqref{eq1} satisfies $\|x(t)\| \leq \beta(\|x(t_0)\|,t-t_0)$ for any $x(t_0) \in D_0$ and $t \geq t_0 \geq 0$, 
then there exists a continuously differentiable function $S(x,t):[0,\infty) \times D_0 \to \mathbb R$ that satisfies the inequality 
$\frac{\partial S(x,t)}{\partial t}+\nabla \{S(x,t)\}^{\rm T} f(x,t) \leq -\alpha(\|x\|)$. Here the function $\alpha(\|\cdot\|)$ is a class $\mathcal K$ functions on $[0,r_0]$.
Next, we consider two cases separately which correspond to the functions $S(x,t)$ and $S^{-1}(x,t)$.
 
1. If $\frac{\partial S(x,t)}{\partial t}+\nabla\{S(x,t)\}^{\rm T}f(x,t)<0$ for any $x \in D_0 \setminus \{0\}$ and $t \geq 0$, 
then
$\frac{\partial S(x,t)}{\partial t}+\frac{1}{|\nabla\{S(x,t)\}|} \nabla\{S(x,t)\}^{\rm T}| \nabla \{S(x,t)\}|f(x,t)<0$. 
Therefore, the following expression holds
\begin{equation*}
\begin{array}{l}
F_1=\int_{V}\frac{\partial S(x,t)}{\partial t} dV
+\oint_{\Gamma} \frac{1}{|\nabla\{S(x,t)\}|} \nabla\{S(x,t)\}^{\rm T}|\nabla\{S(x,t)\}|  f(x,t) d \Gamma<0.
\end{array}
\end{equation*}
 Using Divergence theorem (or Gauss theorem), we get $F_1=\int_{V}\Big[ \frac{\partial S(x,t)}{\partial t}+\nabla \cdot\{|\nabla\{S(x,t)\}|f(x,t)\}\Big]dV<0$.

2. If $\frac{\partial S(x,t)}{\partial t}+\nabla\{S(x,t)\}^{\rm T}f(x,t)<0 $ for any $x \in D_0 \setminus \{0\}$ and $t \geq 0$, 
then $\frac{\partial S^{-1}(x,t)}{\partial t}+\nabla\{S^{-1}(x,t)\}^{\rm T}f(x,t)=
-S^{-2}(x,t)\big[\frac{\partial S(x,t)}{\partial t}+\nabla \cdot \{S (x,t)\}^{\rm T} f(x,t)\big]>0$. 
On the other hand,
$\nabla \{S^{-1}(x,t)\}^{\rm T}f(x,t)
= \frac{1}{|\nabla\{S^{-1}(x,t)\}|} \nabla\{S^{- 1} (x)\}^{\rm T} \times \\ |\nabla\{S^{-1}(x)\}|f(x,t).$
Therefore, the following relation is satisfied
\begin{equation*}
\begin{array}{l}
F_2= \int_{V_{inv}}\frac{\partial S^{-1}(x,t)}{\partial t} dV_{inv}+
\\
\oint_{\Gamma_{inv}}\frac{1}{|\nabla\{S^{-1}(x,t)\}|}
\nabla\{S^{-1}(x,t)\}^{\rm T}|\nabla\{S^{-1}(x,t)\}|f(x,t)d \Gamma_{inv}>0.
\end{array}
\end{equation*}
According to Divergence theorem, we get
$F_2=\int_{V_{inv}}\Big[\frac{\partial S^{-1}(x,t)}{\partial t}+\\ \nabla \cdot\{|\nabla\{S^{-1}(x,t)\}|f(x,t)\}\Big]dV_{inv}>0.$ 
Theorem \ref{Th1} is proved.

$ $

The next theorem extends results of Theorem \ref{Th1} to the case of introducing new auxiliary function in the integrand. It allows to simplify the investigation of stability of system \eqref{eq1}.

\begin{theorem}
\label{Th2}

Let the Jacobian matrix $[\partial f / \partial x]$ be bounded on $D=\{x \in \mathbb R^n: \|x\|<r, r>0\}$ and uniformly in $t$, trajectories of system \eqref{eq1} satisfies 
$\|x(t)\| \leq \beta(\|x(t_0)\|,t-t_0)$ and 
$\|f(x,t)\| \leq c_0 \|x\|^{\gamma}$, $c_0>0$, $\gamma>0$ 
for any $x(t_0) \in D_0$ and $t \geq t_0 \geq 0$, 
where $\beta(\cdot,\cdot)$ is a class $\mathcal{KL}$ function, 
$D_0=\{x \in \mathbb R^n: \|x\|<r_0, r_0>0\}$ and $\beta(r_0,0)<r$.
Then there is a continuously differentiable function 
$S(x,t): [0, \infty) \times D_0 \to \mathbb R$ and the function $\mu(x,t): [0, \infty) \times D_0 \to \mathbb R$ that satisfy $w_1(x) \leq S(x,t) \leq w_2(x)$ and $c_1 \|x\| \leq \mu(x,t) \leq c_2 \|x\|$, $w_1(x)$ and $w_2(x)$ are positive definite functions, $c_1>0$ and $c_2>0$, $|\nabla\{S(x,t)\}| \neq 0$ for any $x \in D_0 \setminus \{0\} $, $t \geq 0$ and at least one of the following conditions holds:

\begin{enumerate}
\item[\textup{(1)}] the function $\frac{\partial S(x,t)}{\partial t}+\nabla \cdot \{\mu(x,t)|\nabla\{S(x,t)\}|f(x,t)\}$ is integrable in the domain $V=\{x \in D_0, t \geq 0: S(x,t) \leq C \}$ and 
$\int_{V} \Big[ \frac{\partial S(x,t)}{\partial t} + \nabla \cdot \{\mu(x,t)|\nabla\{S(x,t)\}|f(x,t)\} \Big] dV <0$ for all $C>0$;

\item[\textup{(2)}] the function $\frac{\partial S^{-1}(x,t)}{\partial t}+\nabla \cdot \{\mu(x,t)|\nabla\{S^{-1}(x,t)\}|f(x,t)\}$ is integrable in the domain $V_{inv}=\{x \in D_0, t \geq 0: S^{-1}(x,t)  \geq C \}$ and 
$\int_{V_{inv}} \Big[\frac{\partial S^{-1}(x,t)}{\partial t}+\nabla \cdot \{\mu(x,t)|\nabla\{S^{-1}(x,t)\}|f(x,t)\}\Big] dV_{inv} >0$ for all $C>0$.
\end{enumerate}

\end{theorem}

\textbf{Proof 2}
According to \cite[Theorem 3.13]{Khalil09}, if the Jacobian matrix $[\partial f / \partial x]$ is bounded on $D=\{x \in \mathbb R^n: \|x\|<r\}$ and uniformly in $t$, trajectories of the system \eqref{eq1} satisfies $\|x(t)\| \leq \beta(\|x(t_0)\|,t-t_0)$ for any $x(t_0) \in D_0$ and $t \geq t_0 \geq 0$, 
then there exists a continuously differentiable function 
$S(x,t):[0,\infty) \times D_0 \to \mathbb R$ 
that satisfies the inequalities 

\begin{equation*}
\begin{array}{l}
S(x,t) \leq \alpha_1(x),
~~
\frac{\partial S(x,t)}{\partial t}+\nabla \{S(x,t)\}^{\rm T} f(t,x) \leq -\alpha_2(x),
~~
\left \|\frac{\partial S(x,t)}{\partial x} \right \| \leq \alpha_3(x).
\end{array}
\end{equation*}
Here $\alpha_1(x)$, $\alpha_2(x)$ and $\alpha_3(x)$ are class $\mathcal{K}$ functions on $[0,r_0]$. If $\alpha_1(x)=c_3\|x\|^{\chi}$, $\chi>0$, then $\alpha_2(x)=c_4 \|x\|^{\chi+\gamma-1}$ and  $\alpha_3(x)=c_5 \|x\|^{\chi-1}$
Consider the following relations

\begin{equation*}
\begin{array}{l}
\frac{\partial S(x,t)}{\partial t}+\nabla\{S(x,t)\}^{\rm T}\mu(x,t)f(x,t) 

\\
= 
\frac{\partial S(x,t)}{\partial t}+\nabla\{S(x,t)\}^{\rm T}f(x,t)+\nabla\{S(x,t)\}^{\rm T}(\mu(x,t)-1)f(x,t) 

\\
\leq
\frac{\partial S(x,t)}{\partial t}+\nabla\{S(x,t)\}^{\rm T}f(x,t)
+\|\nabla\{S(x,t)\}\| |\mu(x,t)| \|f(x,t)\|

\\
\leq
 -(c_4-c_0 c_2 c_5 \|x\|)\|x\|^{\chi+\gamma-1}.
\end{array}
\end{equation*}
Choosing $r=\frac{c_4}{c_0 c_2 c_5}$ and $r_0<r$, one gets $\frac{\partial S(x,t)}{\partial t}+\nabla\{S(x,t)\}^{\rm T}\mu(x,t)f(x,t) \leq 0$.
Next, we consider two cases separately which correspond to the functions $S(x,t)$ and $S^{-1}(x,t)$.

1. Since
$\frac{\partial S(x,t)}{\partial t}+\nabla\{S(x,t)\}^{\rm T}\mu(x,t)f(x,t)<0$  
for any $x \in D_0 \setminus \{0\}$ and $t \geq 0$, then
\begin{equation*}
\begin{array}{l}
F_1=\int_{V}\frac{\partial S(x,t)}{\partial t} dV
+\oint_{\Gamma}\frac{1}{|\nabla\{S(x,t)\}|}\nabla\{S(x,t)\}^{\rm T}\mu(x,t)|\nabla\{S(x,t)\}|f(x,t)d \Gamma <0.
\end{array}
\end{equation*}
Using Divergence theorem, we get $F_1=\int_{V}\Big[ \frac{\partial S(x,t)}{\partial t}+\nabla \cdot\{|\nabla\{S(x,t)\}|\mu(x,t) \\ f(x,t)\}\Big]dV<0$.

2. Since
$\frac{\partial S(x,t)}{\partial t}+\nabla\{S(x,t)\}^{\rm T}\mu(x,t)f(x,t)<0$ 
for any $x \in D_0 \setminus \{0\}$ and $t \geq 0$, 
then $\frac{\partial S^{-1}(x,t)}{\partial t}+\nabla\{S^{-1}(x,t)\}^{\rm T}\mu(x,t)f(x,t) = -S^{-2}(x,t)\Big[\frac{\partial S(x,t)}{\partial t}+\nabla\{S(x,t)\}^{\rm T}\mu(x,t)f(x,t)\Big] > 0$.
On the other hand,
$\nabla \{S^{-1}(x,t)\}^{\rm T}\mu(x,t)f(x,t) \\
= \frac{1}{|\nabla \{S^{-1}(x,t)\}|} \nabla\{S^{- 1} (x)\}^{\rm T} |\nabla\{S^{- 1}(x)\}|\mu(x,t) f(x,t).$
Therefore, the following relation is satisfied
\begin{equation*}
\begin{array}{l}
F_2=\int_{V_{inv}}\frac{\partial S^{-1}(x,t)}{\partial t} dV_{inv}
\\
+\oint_{\Gamma_{inv}}\frac{1}{|\nabla\{S^{-1}(x,t)\}|}
\nabla\{S^{-1}(x,t)\}^{\rm T} 
|\nabla\{S^{-1}(x,t)\}|\mu(x,t)
f(x,t) d\Gamma_{inv}>0.
\end{array}
\end{equation*}
Considering Divergence theorem, we get
$F_2=\int_{V_{inv}}\Big[\frac{\partial S^{-1}(x,t)}{\partial t}+\nabla \cdot\{|\nabla\{S^{-1}(x,t)\} \\ | \mu(x,t) f(x,t)\}\Big] dV_{inv}>0.$ 
Theorem \ref{Th2} is proved.

\textbf{Remark 1}
\label{rem2a}
There are various physical interpretations of the integral conditions in Theorem \ref{Th2}. 
Rewriting the integral inequalities in Theorem \ref{Th2} as
$\int_{V} \Big[\frac{\partial S(x,t)}{\partial t}+\nabla \cdot \{\mu(x,t)|\nabla\{S(x,t)\}|f(x,t)\}\Big] dV = -\Sigma$ or $\int_{V_{inv}} \Big[\frac{\partial S^{-1}(x,t)}{\partial t}+\nabla \cdot \{\mu(x,t)|\nabla\{S^{-1}(x,t)\}|f(x,t)\}\Big] dV_{inv} = \Sigma$, $\Sigma \geq 0$, one gets the integral forms of continuity equation with the sources (the flux is directed inward) located in the equilibrium points in the domain $V$ (or $V_{inf}$), see \cite{Arnold14}. 

\begin{enumerate}
\item[(i)] Choosing $\mu(x,t)$ such that $|\nabla\{S(x,t)\}|\mu(x,t)=S(x,t)$ or $|\nabla\{S^{-1}(x,t)\}| \\ \times \mu(x,t)=S^{-1}(x,t)$, one has the continuity equation in fluid dynamics \cite{Pedlosky79}, where $S$ is fluid density and $f$ is the flow velocity of a vector field. 
\item[(ii)]  In electromagnetic theory \cite{Griffiths17}, $\mu(x,t)|\nabla\{S(x,t)\}|f(x,t)$ or $\mu(x,t)| \times \\ \nabla\{S^{-1}(x,t)\}|f(x,t)$ means the current density and $S$ is the charge density. 
\item[(iii)] Due to conservation of energy \cite{Pedlosky79}, $\mu(x,t)|\nabla\{S(x,t)\}|f(x,t)$ or $\mu(x,t)| \times \\ \nabla\{S^{-1}(x,t)\}|f(x,t)$ is the vector energy flux and $S$ is local energy density. 
\item[(iv)] In quantum mechanics \cite{McMahon13}, $S$ is the probability density function and $\mu(x,t)|\nabla\{S(x,t)\}|f(x,t)$ or $\mu(x,t)|\nabla\{S^{-1}(x,t)\}|f(x,t)$ is the probability current.
\end{enumerate}

\textbf{Remark 2}
\label{rem3a}
If $\frac{\partial S(x,t)}{\partial t}+\nabla \cdot \{\mu(x,t)|\nabla\{S(x,t)\}|f(x,t)\}$ or $\frac{\partial S^{-1}(x,t)}{\partial t}+\nabla \cdot \{\mu(x,t)|\nabla\{S^{-1}(x,t)\}|f(x,t)\}$ 
are integrable and 
$\frac{\partial S(x,t)}{\partial t}+\nabla \cdot \{\mu(x,t)|\nabla\{S(x,t)\} \\ |f(x,t)\}= -\sigma$ or 
$\frac{\partial S^{-1}(x,t)}{\partial t}+\nabla \cdot \{\mu(x,t)|\nabla\{S^{-1}(x,t)\}|f(x,t)\} = 
\sigma$, $\sigma \geq 0$ holds for any 
$x \in D_0 \setminus \{0\}$ anf $t \geq 0$, 
then the corresponding integral relations in Theorem \ref{Th2} are satisfied. 
According to \cite{Pedlosky79,Griffiths17,McMahon13,Arnold14} and Remark \ref{rem2a}, one gets the appropriate differential forms of continuity equations with the sources (the flux is directed inward) in the domain $V$ (or $V_{inf}$).

\textbf{Remark 3}
\label{rem3a}
According to \cite{Rantzer01}, if the relation $\frac{\partial S^{-1}(x,t)}{\partial t}+\nabla \cdot \{S^{-1}(x,t) f(x,t)\}>0$ holds, then only almost all solutions of \eqref{eq1} tend to equlibrium. Thus, the results of \cite{Rantzer01} are special case in Theorem \ref{Th2} for $|\nabla\{S^{-1}(x,t)\}|\mu(x,t)=S^{-1}(x,t)$.

Now let us formulate a sufficient condition for stability of \eqref{eq1}.

\begin{theorem}
\label{Th2a}
Let $S(x,t):[0,\infty) \times D \to \mathbb R$ be a continuously differentiable function such that $w_1(x) \leq S(x,t) \leq w_2(x)$ for any $t \geq 0$ and $x \in D$, where $w_1(x)$ and $w_2(x)$ are positive definite continuously differentiable functions. Then $x=0$ is uniformly stable if at least one of the following conditions holds:

\begin{enumerate}
\item[\textup{(1)}] $\frac{\partial S(x,t)}{\partial t}+\nabla \cdot \{S(x,t) f(x,t)\} \leq S(x,t)\nabla \cdot\{f(x,t)\}$ for any $t \geq 0$ and $x \in D$;

\item[\textup{(2)}] $\frac{\partial S^{-1}(x,t)}{\partial t}+\nabla \cdot \{S^{-1}(x,t)f(x,t)\} \geq 0$ and $\nabla \cdot\{f(x,t)\} \leq 0$ for any $t \geq 0$ and $x \in D$;

\item[\textup{(3)}] 
$2\frac{\partial S(x,t)}{\partial t}+\nabla \cdot\{S(x,t)f (x)\} \leq 0$ and $\nabla \cdot\{S^{-1}(x)f(x,t)\} \geq 0$  
 for any $t \geq 0$ and $x \in D$.

\end{enumerate}

If inequalities are strict in the cases \textup{(1)-(3)}, then $x=0$ is uniformly asymptotically stable.
\end{theorem}

\textbf{Proof 3}
Consider the proof for each case separately. The proof of asymptotic stability is omitted because it is similar to the proof of stability, but taking into account the sign of a strict inequality.

1. From the relation 
$\nabla \cdot \{S(x,t) f(x,t)\}-\nabla \cdot\{f(x,t)\} S(x,t) = \nabla\{S(x,t)\}^{\rm T} \times \\ f(x,t)$ 
implies that if 
$\frac{\partial S(x,t)}{\partial t}+\nabla \cdot\ \{S(x,t) f(x,t)\} \leq \nabla \cdot\{f(x,t)\} S(x,t)$, 
then 
$\frac{\partial S(x,t)}{\partial t}+\nabla\{S(x,t)\}^{\rm T} f(x,t) \leq  0$ for any $x \in D \setminus \{0\}$ and $t \geq 0$. 
Therefore, according to Lyapunov theorem \cite{Khalil09}, system \eqref{eq1} is stable.

2. From the expression
$S^{-2}(x,t)\nabla\{S(x,t)\}^{\rm T} f(x,t)=S^{-1}(x,t)\nabla \cdot\{f(x,t)\} \\ -\nabla \cdot\{S^{-1}(x,t)f(x,t)\}$
follows that if 
$\frac{\partial S^{-1}(x,t)}{\partial t}+\nabla \cdot\{S^{-1}(x,t)f(x,t)\} \geq 0$ and $\nabla \cdot \{f(x,t)\} \leq 0$, 
then 
$\frac{\partial S(x,t)}{\partial t}+\nabla\{S(x,t)\}^{\rm T} f(x,t) \leq 0$ for any $x \in D \setminus \{0\}$ and $t \geq 0$. 

3. Condition 3 is a combination of conditions 1 and 2. 
Introduce $\beta(x,t) \geq 1$ for any $x \in D \setminus \{0\}$ and $t \geq 0$. 
Summing $\beta(x,t) \nabla \{S(x,t)\}^{\rm T} f(x,t)=\beta(x,t) S(x,t)\nabla \cdot \{f(x,t)\}-\beta(x,t) S^{2}(x,t) \nabla \cdot\ \{S^{-1}(x,t)f(x,t)\}$ 
and 
$\nabla\{S(x,t)\}^{\rm T}  \\ \times f(x,t)=
\nabla \cdot\{S(x,t)f(x,t)\}-\nabla \cdot\{f(x,t)\}S(x,t)$, 
we get 

\begin{equation*}
\begin{array}{l}
(1+\beta(x,t)) \nabla\{S(x,t)\}^{\rm T}f(x,t)=\nabla \cdot\{S(x,t)f (x,t)\}-
\\
\beta(x,t)S^{2}(x,t) \nabla \cdot\{S^{-1}(x,t)f(x,t)\}+(\beta(x,t)-1)S(x,t)\nabla \cdot\{f(x,t)\}.
\end{array}
\end{equation*}
If
\begin{equation*}
\label{common}
\begin{array}{l}
\nabla \cdot\{S(x,t)f (x,t)\}+(\beta(x,t)-1)S(x,t)\nabla \cdot\{f (x,t)\} \leq 
\\
\beta(x,t)S^{2}(x,t) \nabla \cdot\{S^{-1}(x,t)f(x,t)\},
\end{array}
\end{equation*}
then $\nabla\{S(x,t)\}^{\rm T}f(x,t) \leq 0$. Let $\beta(x,t)=1$. If $2 \frac{\partial S(x,t)}{\partial t}+\nabla \cdot\{S(x,t)f (x)\} \leq 0$ and $\nabla \cdot\{S^{-1}(x)f(x,t)\} \geq 0$, then $\nabla\{S(x,t)\}^{\rm T}f(x,t) \leq 0$ holds for any $x \in D \setminus \{0\}$ and $t \geq 0$. 
Theorem \ref{Th2a} is proved.

$ $

It is noted in Introduction that the result of \cite{Zaremba54,Shestakov78,Zhukov99} is applicable only to second-order autonomous systems. Next, we consider an illustration of the proposed results for third-order nonautonomous systems and compare the results with ones from \cite{Rantzer01}.

\textit{Example 1}.
Consider the system

\begin{equation}
\label{eq6}
\begin{array}{l} 
\dot{x}_1=g(t)x_2-\varphi_1(t) x_1 x_3^2,
\\
\dot{x}_2=-x_1-\varphi_2(t) x_2x_3^2,
\\
\dot{x}_3=-\varphi_3(t) x_3^3,
\end{array}
\end{equation} 
which has an equilibrium point $(0,0,0)$. The function $g(t)>0$ is continuously differentiable and bounded, $\dot{g}(t)<0$ for any $t \geq 0$. The functions $\varphi_1(t)>0$, $\varphi_2(t)>0$, and $\varphi_3(t)>0$ are continuous and bounded for any $t \geq 0$.

Choose $S(x,t)=(x_1^2+g(t)x_2^2+x_3^2)^{\alpha}$, where $\alpha$  is a positive integer. 
Verify the conditions of Theorem \ref{Th2}, where $\mu(x,t)$ is chosen such that $|\nabla\{S(x,t)\}|\mu(x,t) \\ = S(x,t)$ or $|\nabla\{S^{-1}(x,t)\}|\mu(x,t)=S^{-1}(x,t)$.
The condition 
\begin{equation*}
\begin{array}{l} 
\frac{\partial S(x,t)}{\partial t}+\nabla \cdot\{S(x,t)f(x,t)\}
\\
=-2\alpha (x_1^2+g(t)x_2^2+x_3^2)^{\alpha-1}[x_3^2(\varphi_1 x_1^2+g\varphi_2 x_2^2+\varphi_3 x_3^2)-0.5\dot{g}x_2^2]<0
\end{array}
\end{equation*}
holds for any $\alpha$, $x_2 \neq 0$, $x_3 \neq 0$, and $t \geq 0$. 
The relation 
\begin{equation*}
\begin{array}{l} 
\frac{\partial S^{-1}(x,t)}{\partial d}+\nabla \cdot\{S^{-1}(x)f (x)\}
=
(x_1^2+g(t)x_2^2+x_3^2)^{-\alpha-1} \times
\\
\Big(x_3^2\big[(2\alpha\varphi_1-\varphi_0)x_1^2+g(2\alpha \varphi_2-\varphi_0)x_2^2+
(2\alpha\varphi_3-\varphi_0)x_3^2\big]-\alpha \dot{g}x_2^2\Big)>0
\end{array}
\end{equation*}
 is satisfied for any $x_2 \neq 0$, $x_3 \neq 0$, $t \geq 0$, and 
 $\alpha \geq 0.5\sup\limits_{t}\left\{
 \frac{\varphi_0(t)}{\varphi_1(t)},
 \frac{\varphi_0(t)}{\varphi_2(t)},
 \frac{\varphi_0(t)}{\varphi_3(t)}\right\}$.  
Therefore, the conditions of Theorem \ref{Th2} are fulfilled. 
Since the function $\nabla \cdot\{S^{-1}(x)f(x,t)\} $ is integrable in 
$\{x \in \mathbb R^n: |x| \geq 1 \}$, 
then the conditions of Theorem 1 and Corollary 1 in \cite{Rantzer01} 
(convergence of almost all solutions of \eqref{eq6}) are satisfied too.

Now let us verify the conditions of Theorem \ref{Th2a}. 
The relation 
\begin{equation*}
\begin{array}{l} 
\frac{\partial S(x,t)}{\partial t}+\nabla \cdot\{S(x,t)f(x,t)\}-S(x,t)\nabla \cdot\{f(x,t)\}
\\
=
-2\alpha (x_1^2+g(t)x_2^2+x_3^2)^{\alpha-1}[x_3^2(\varphi_1 x_1^2+g\varphi_2 x_2^2+\varphi_3 x_3^2)-0.5\dot{g}x_2^2]<0
\end{array}
\end{equation*}
 holds for any $\alpha$, $x_2 \neq 0$, $x_3 \neq 0$, and $t \geq 0$.
In turn, $\nabla \cdot\{f(x,t)\}=-\varphi_0 x_3^2<0$, $\varphi_0
=\varphi_1+\varphi_2+3\varphi_3$ 
and the condition 
\begin{equation*}
\begin{array}{l} 
\frac{\partial S^{-1}(x,t)}{\partial t}+\nabla \cdot\{S^{-1}(x)f(x,t)\}
=(x_1^2+g(t)x_2^2+x_3^2)^{-\alpha-1} \times
\\
\Big(x_3^2\Big[(2\alpha\varphi_1-\varphi_0)x_1^2+g(2\alpha \varphi_2-\varphi_0)x_2^2+(2\alpha\varphi_3-\varphi_0)x_3^2\Big]-\alpha \dot{g}x_2^2\Big)>0
\end{array}
\end{equation*}
 holds for any $x_2 \neq 0$, $x_3 \neq 0$, $t \geq 0$, and 
 $\alpha \geq 0.5\sup\limits_{t}\left\{
 \frac{\varphi_0(t)}{\varphi_1(t)},
 \frac{\varphi_0(t)}{\varphi_2(t)},
 \frac{\varphi_0(t)}{\varphi_3(t)}\right\}$. 
The raltions 
\begin{equation*}
\begin{array}{l} 
2\frac{\partial S(x,t)}{\partial t}+
\nabla \cdot\{S(x,t)f(x,t)\}=
\\
-2\alpha (x_1^2+g(t)x_2^2+x_3^2)^{\alpha-1}[x_3^2(\varphi_1 x_1^2+g\varphi_2 x_2^2+\varphi_3 x_3^2)-\dot{g}x_2^2]<0
\end{array}
\end{equation*}
and 
\begin{equation*}
\begin{array}{l} 
\nabla \cdot\{S^{-1}(x)f(x,t)\}
=(x_1^2+g(t)x_2^2+x_3^2)^{-\alpha-1} \times 
\\
\Big(x_3^2\Big[(2\alpha\varphi_1-\varphi_0)x_1^2+g(2\alpha \varphi_2-\varphi_0)x_2^2+
(2\alpha\varphi_3-\varphi_0)x_3^2\Big]\Big)>0
\end{array}
\end{equation*}
 are satisfied for any $x_2 \neq 0$, $x_3 \neq 0$, $t \geq 0$, and 
 $\alpha \geq 0.5\sup\limits_{t}\left\{
 \frac{\varphi_0(t)}{\varphi_1(t)},
 \frac{\varphi_0(t)}{\varphi_2(t)},
 \frac{\varphi_0(t)}{\varphi_3(t)}\right\}$.   
All three cases gave the same results. 
Therefore, $x=0$ is uniformly asymptotically stable with any initial conditions when $x_3(0) \neq 0$. 
If the initial conditions contain $x_2 (0)=x_3 (0)=0$, then $x=0$ is uniformly stable. 

\begin{figure}[h]
\begin{minipage}[h]{0.5\linewidth}
\center{\includegraphics[width=1\linewidth]{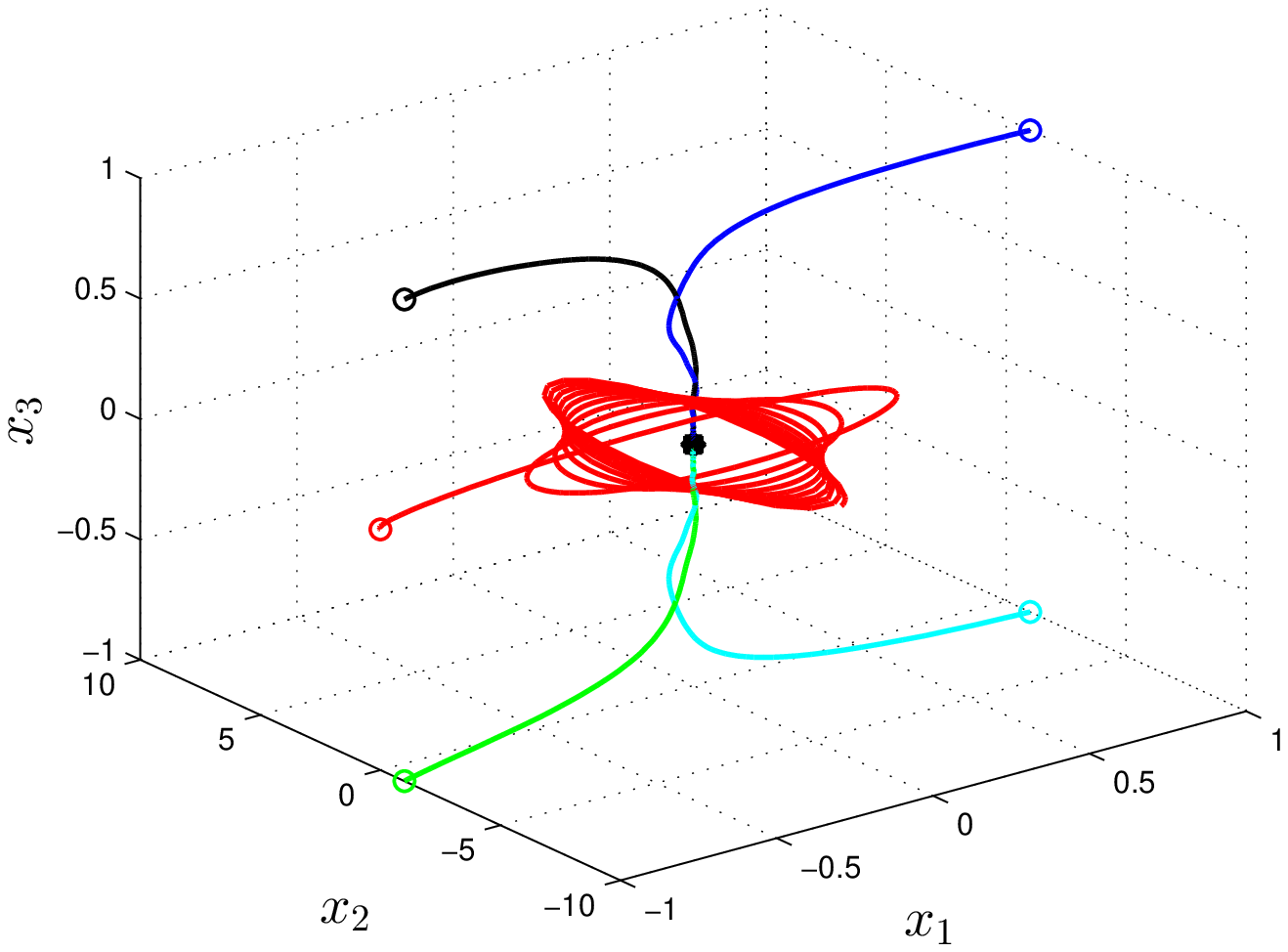}}
\end{minipage}
\hfill
\begin{minipage}[h]{0.5\linewidth}
\center{\includegraphics[width=1\linewidth]{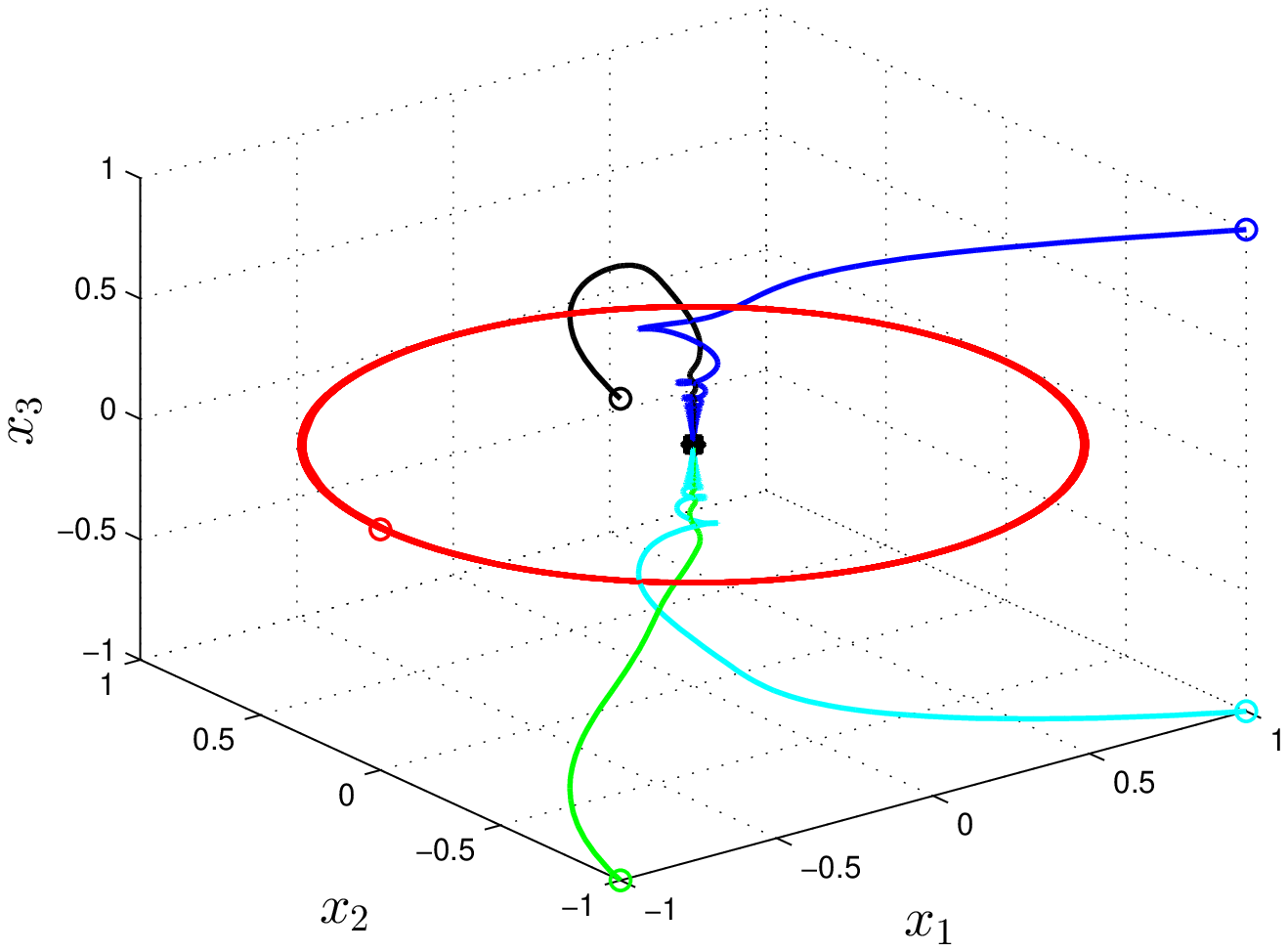}}
\end{minipage}
\caption{Phase trajectories of system \eqref{eq6} for $g=\frac{1}{t+1}$ (left picture) and $g=1$ (right picture).}
\label{Fig5}
\end{figure}

The phase trajectories of \eqref{eq6} are shown in Fig.~\ref{Fig5} for 
$\varphi_1=2+\sin(2t)$, $\varphi_2=1.5+\cos(3t)$,  $\varphi_3=\frac{t}{t+1}$ and $g=\frac{1}{t+1}$ (left picture) or $g=1$ (right picture). In Fig.~\ref{Fig5} the cycles are obtained for the initial conditions with $x_2 (0)=x_3 (0)=0$, the converging to zero curves are obtained for $x_2 (0) \neq 0$ and $x_3 (0) \neq 0$.

As a result, the proposed Theorem \ref{Th2} and Corollary 1 from \cite{Rantzer01} give positive answers about the possible stability of \eqref{eq6}. The conditions of Theorem \ref{Th2a} have established that $x=0$ is uniformly asymptotically stable or uniformly stable depending on the values of the initial conditions.

\begin{figure}[h!]
\center{\includegraphics[width=0.7\linewidth]{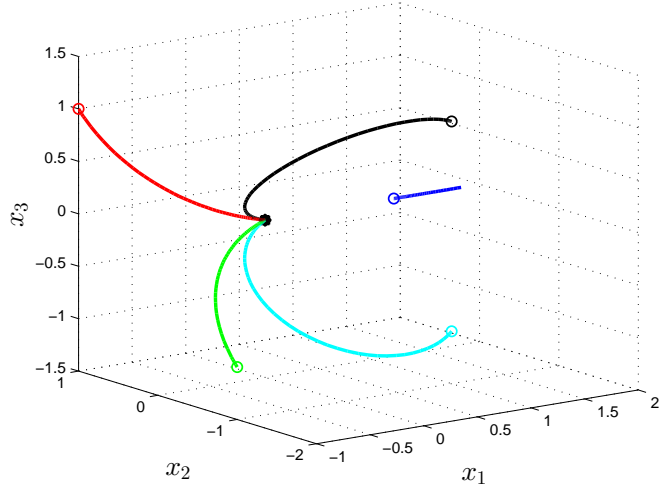}}
\caption{Phase trajectories of system \eqref{eq04aa} with two equilibrium points.}
\label{Ust_Rantz_3rd}
\end{figure}

\textit{Example 2}.
Consider the system

\begin{equation}
\label{eq04aa}
\begin{array}{l} 
\dot{x}_1=-x_1+x_1^2-\frac{1}{g(t)}x_2^2-x_3^2,
\\
\dot{x}_2=-x_2+2x_1 x_2,
\\
\dot{x}_3=-x_3+2x_1 x_3,
\end{array}
\end{equation} 
which has two equilibrium points $(0,0,0)$ and $(1,0,0)$. 
The function $g(t)>0$ is continuously differentiable and bounded, $\dot{g}(t)<0$ for any $t \geq 0$.
All trajectories of the system converge to the point $(0,0,0)$, 
except those that start on the semi-axis $x_1 \geq 1$, $x_2=0 $ and $x_3=0$ (see Fig.~\ref{Ust_Rantz_3rd} for $g(t)=\frac{1}{t+1}$). 
Let $S(x,t)=(gx_1^2+x_2^2+gx_3^2)^{\alpha}$, $\alpha$ is a positive integer and $\mu(x,t)$ in Theorem \ref{Th2} is chosen such that $|\nabla\{S(x,t)\}|\mu(x,t)=S(x,t)$ or $|\nabla\{S^{-1}(x,t)\}|\mu(x,t)=S^{-1}(x,t)$. 
Then inequality
\begin{equation*}
\begin{array}{l} 
\frac{\partial S^{-1}(x,t)}{\partial t}+\nabla \cdot\{S^{-1}(x)f(x,t)\}=
-\alpha(gx_1^2+x_2^2+gx_3^2)^{-\alpha-1}\dot{g} (x_1^2+x_3^2)
\\
+(gx_1^2+x_2^2+gx_3^2)^{-\alpha}[2\alpha-3+2x_1(3-\alpha)]>0
\end{array}
\end{equation*}
holds for $\alpha=3$. The function $\nabla \cdot \{f(x,t)\}=-3+6x_1$ does not satisfy the condition $\nabla \cdot\{f(x,t)\} \leq 0$ for $x_1>0.5$. 
The relations $\frac{\partial S(x,t)}{\partial d}+\nabla \cdot\{S(x,t)f(x,t)\} \leq S(x,t) \nabla \cdot\{f(x,t)\}$ and $2\frac{\partial S(x,t)}{\partial t}+\nabla \cdot\{S(x,t)f(x,t)\} \leq 0$ from Theorem \ref{Th2a} are not satisfied too. 
As a result, the conditions of the proposed Theorem \ref{Th2} (and the conditions of Corollary 1 in \cite{Rantzer01}) are fulfilled in this example, but the conditions of Theorem \ref{Th2a} are not satisfied.

\textit{Example 3.}
Consider the system

\begin{equation}
\label{eq6aa}
\begin{array}{l} 
\dot{x}_1=-4 x_1 x_2^2 - \varphi_1(t) x_1^3,
\\
\dot{x}_2=g_1(t) x_1^2 x_2-\varphi_2(t) x_2^3-g_2(t) x_2 x_3^2,
\\
\dot{x}_3=-\varphi_3(t) x_3^3+8x_2^2x_3
\end{array}
\end{equation}
with equilibrium point $(0,0,0)$.  
The functions $g_1(t)>0$ and $g_2(t)>0$ are continuously differentiable and bounded, $\dot{g}_1(t)<0$ and $\dot{g}_2(t)<0$ for any $t \geq 0$. 
The functions $\varphi_1(t)>0$, $\varphi_2(t)>0$, and $\varphi_3(t)>0$ are continuous and bounded for any $t \geq 0$.

Choose $S(x,t)=(\frac{1}{8}g_1x_1^2+\frac{1}{2}x_2^2 + \frac{1}{16} g_2 x_3^2)^{\alpha}$, $\alpha$  is a positive integer. 
Verify the conditions of Theorem \ref{Th2a}. 
The relation
\begin{equation*}
\begin{array}{l}
\frac{\partial S(x,t)}{\partial t}+\nabla \cdot\{S(x,t)f (x)\}-S(x,t)\nabla \cdot\{f(x,t)\}=
\\
-\alpha \left (\frac{1}{8}g_1x_1^2+\frac{1}{2}x_2^2 + \frac{1}{16} g_2 x_3^2 \right)^{\alpha-1}
\Big[\frac{1}{4}g_1\varphi_1 x_1^4+\varphi_2 x_2^4 + \frac{1}{8} g_2 x_3^4 -
\frac{1}{8}\dot{g}_1x_1^2 - \frac{1}{16} \dot{g}_2 x_3^2\Big]
\end{array}
\end{equation*}
holds for any  $\alpha$ and $x \neq 0$.
The function $\nabla \cdot \{f(x,t)\}=(g_1-3\varphi_1)x_1^2+(4-3\varphi_2)x_2^2-(3\varphi_3+g_2)x_3^2$ is not negative definite for $g_1 > 3\varphi_1$ and/or $\varphi_2<\frac{4}{3}$. 
Thus, Proposition 2 with taking into account Corollary 1 from \cite{Rantzer01} and the second case of Theorem \ref{Th2a} are not satisfied.
The conditions 
$2\frac{\partial S(x,t)}{\partial t}+\nabla \cdot\{S(x,t)f(x,t)\}<0$ and $\nabla \cdot\{S^{-1}(x,t)f(x,t)\}>0$ hold for any $\alpha$, $x \neq 0$, $\varphi_2>\frac{4}{2\alpha+3}$, and $g_1<(2\alpha+3)\varphi_1$.

\begin{figure}[h!]
\center{\includegraphics[width=0.7\linewidth]{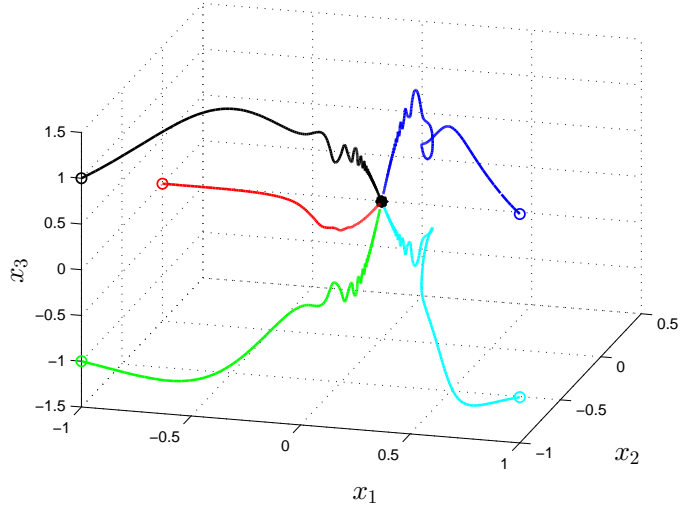}}
\caption{Phase trajectories of system \eqref{eq6aa}.}
\label{Fig5aa}
\end{figure}

Fig.~\ref{Fig5aa} shows the phase trajectories for $\varphi_1(t)=2+\sin(t)$, $\varphi_2(t)=1.5+\cos(3t)$, $\varphi_1(t)=1+0.5\cos(2t)$, $g_1(t)=\frac{t}{t+1}$, and $g_2(t)=\frac{1}{t+1}$. 
Taking into account Theorem \ref{Th2}, the condition $\int_V [\frac{\partial S(x,t)}{\partial t}+\nabla \cdot \{S(x,t)f(x,t)\}]dV<0$ holds for any $C$ and $\alpha$.
The other conditions of Theorem \ref{Th2} and Corollary 1 in \cite{Rantzer01} are not satisfied.

As a result, the conditions of Theorem \ref{Th2} and Theorem \ref{Th2a} are satisfied for system \eqref{eq6aa}. Thus, $(0,0,0)$ is an uniformly asymptotically stable equilibrium point. The conditions of Corollary 1 from \cite{Rantzer01} are not sutisfied and we cannot conclude about convergence of almost all solutions of \eqref{eq6aa} to $(0,0,0)$.

\textit{Example 4.}
Consider the linear system $\dot{x}=A(t)x$, $x \in \mathbb R^n$. 
Let $S(x,t)=(x^{\rm T}P(t)x)^{\alpha}$, where $\alpha>0$ and $P(t)=P^{\rm T}(t)>0$. 
According to the case (3) of Theorem \ref{Th2a}, the relations 
\begin{equation*}
\begin{array}{l}
2\frac{\partial S(x,t)}{\partial t}+\nabla \cdot\{S(x,t)f (x)\}=\alpha(x^{\rm T}Px)^{\alpha-1} \times
\\
 x^{\rm T}[2\dot{P}(t)+A(t)^{\rm T}P(t)+P(t)A(t)+\frac{1}{\alpha} trace(A(t))P(t)]x \leq 0
\end{array}
\end{equation*}
and
\begin{equation*}
\begin{array}{l}\nabla \cdot\{S^{-1}(x)f(x,t)\} =-\alpha(x^{\rm T}Px)^{-\alpha-1} \times
\\
x^{\rm T}[A(t)^{\rm T}P(t)+P(t)A(t)-\frac{1}{\alpha} trace(A(t))P(t)]x \geq 0
\end{array}
\end{equation*}
 hold if 
\begin{equation*}
\begin{array}{l}
2\dot{P}(t)+A(t)^{\rm T}P(t)+P(t)A(t)+\frac{1}{\alpha} trace(A(t))P(t)<0
\end{array}
\end{equation*}
 and 
 \begin{equation*}
\begin{array}{l}
A(t)^{\rm T}P(t)+P(t)A(t)-\frac{1}{\alpha} trace(A(t))P(t) < 0
\end{array}
\end{equation*}
 are simultaneously satisfied. It is obvious, that the sum of these inequalities give nonstationary Lyapunov inequality $\dot{P}(t)+A(t)^{\rm T}P(t)+P(t)A(t)<0$.

\section{Control law design}
\label{Sec3}

Consider a nonautonomous system in the form

\begin{equation}
\label{eq9}
\begin{array}{l} 
\dot{x}=\xi(x,t)+g(x,t)u(x,t),
\end{array}
\end{equation} 
where $x \in \mathbb R^n$ and $u \in \mathbb R^m$ is the control signal. The functions $\xi(x,t): [0, \infty) \times D \to \mathbb R^{n}$, $g(x,t): [0, \infty) \times D \to \mathbb R^{n \times m}$ and $u(x,t): [0, \infty) \times D \to \mathbb R^{m}$ are piecewise continuous in $t$ and continuously differentiable in $x$ on $[0, \infty) \times D$. 
The open set $D \subset \mathbb R^{n}$ contains the origin $x=0$ and $\xi(t,0)=0$, $g(t,0)=0$, $u(t,0)=0$
 for any $t \geq 0$. System \eqref{eq9} is controllable in $D$ for any $t \geq 0$.

\begin{theorem}
\label{Th6}
Let $S(x,t):[0,\infty) \times D \to \mathbb R$ be a continuously differentiable function such that $w_1(x) \leq S(x,t) \leq w_2(x)$ for any $t \geq 0$ and $x \in D$, where $w_1(x)$ and $w_2(x)$ are positive definite continuously differentiable functions. Then the equilibrium point $x=0$ of the closed-loop system is uniformly stable if the control law $u(x,t)$ is chosen such that at least one of the following conditions holds:

\begin{enumerate}
\item[\textup{(1)}] $\frac{\partial S(x,t)}{\partial t}+\nabla \cdot\{S(x,t)(\xi(x,t)+g(x,t)u(x,t))\} \leq S(x,t) \nabla \cdot\{\xi(x,t)+g(x,t)u(x,t)\}$ for any $x \in D \setminus \{0\}$ and $t \geq 0$;

\item[\textup{(2)}] $\frac{\partial S^{-1}(x,t)}{\partial t}+\nabla \cdot\{S^{-1}(x,t)(\xi(x,t)+g(x,t)u(x,t))\} \geq 0$ and $\nabla \cdot\{\xi(x,t)+g(x,t)u(x,t)\} \leq 0$ for any $x \in D \setminus \{0\}$ and $t \geq 0$;

\item[\textup{(3)}] $2 \frac{\partial S(x,t)}{\partial t}+\nabla \cdot\{S(x,t)(\xi(x,t)+g(x,t)u(x,t))\} \leq 0$ and 
$\nabla \cdot\{S^{-1}(x,t)(\xi(x,t)+g(x,t)u(x,t))\} \geq 0$ for any $x \in D \setminus \{0\}$  and $t \geq 0$.
\end{enumerate}

If the control law $u(x,t)$ is chosen such that in the cases \textup{(1)-(3)} the inequalities are strict, then $x=0$ is uniformly asymptotically stable.
\end{theorem}

Since system \eqref{eq9} is controllable in $D$, the proof of Theorem \ref{Th6}  is similar to the proof of Theorem \ref{Th2a} (denoting by $f(x,t)=\xi(x,t)+g(x,t)u(x,t)$).

If the control law design is based on the method of Lyapunov functions, then it is required to solve the algebraic inequality $\frac{\partial V}{\partial t}+\nabla\{V\}(f+gu)<0$ \textit{w.r.t.} $u$.  According to Theorem \ref{Th6}, the control law is chosen from the feasibility of differential inequality. This gives new opportunities for the control law design.

\begin{figure}[h]
\begin{minipage}[h]{0.5\linewidth}
\center{\includegraphics[width=1\linewidth]{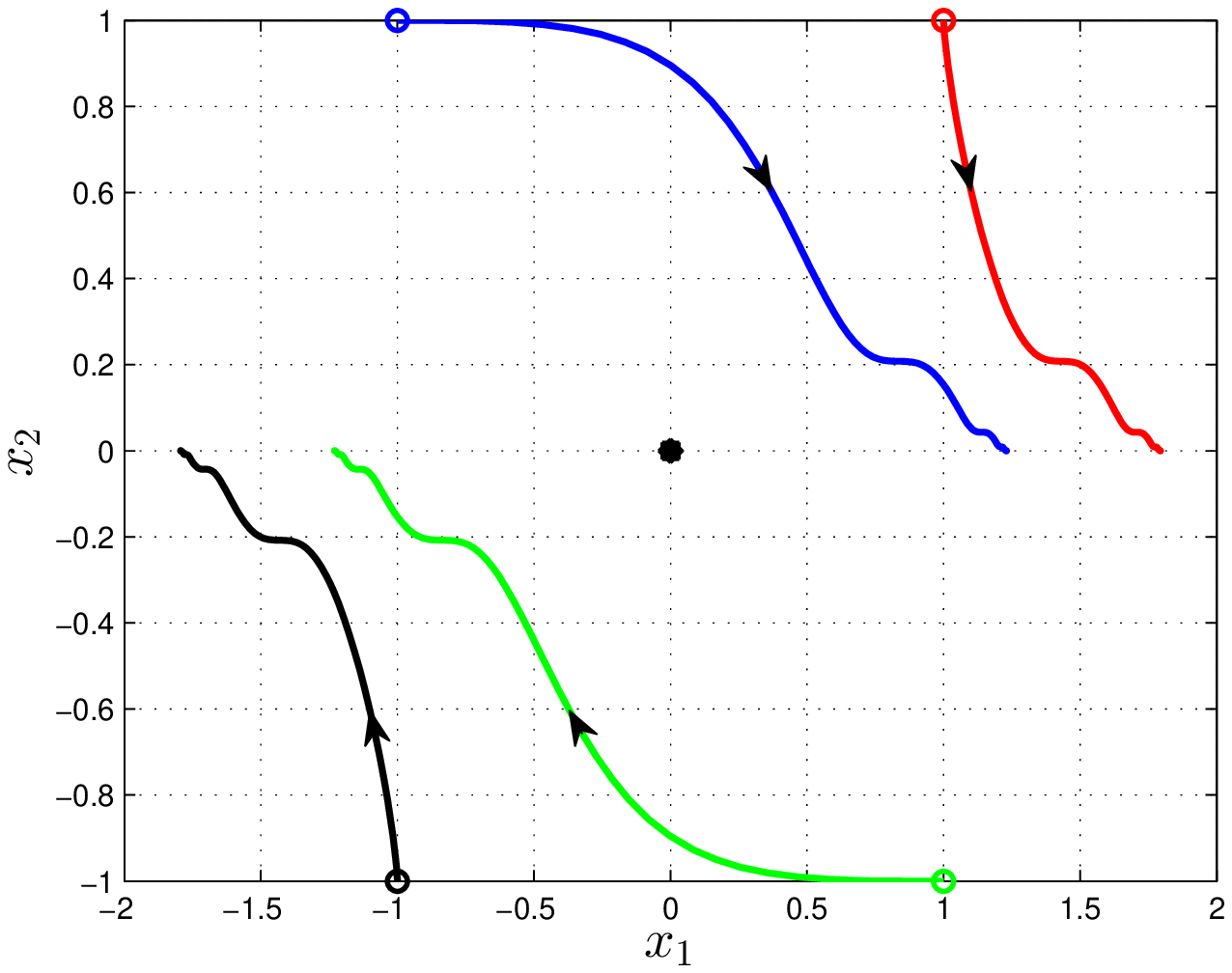}} \textit{a}
\end{minipage}
\hfill
\begin{minipage}[h]{0.5\linewidth}
\center{\includegraphics[width=1\linewidth]{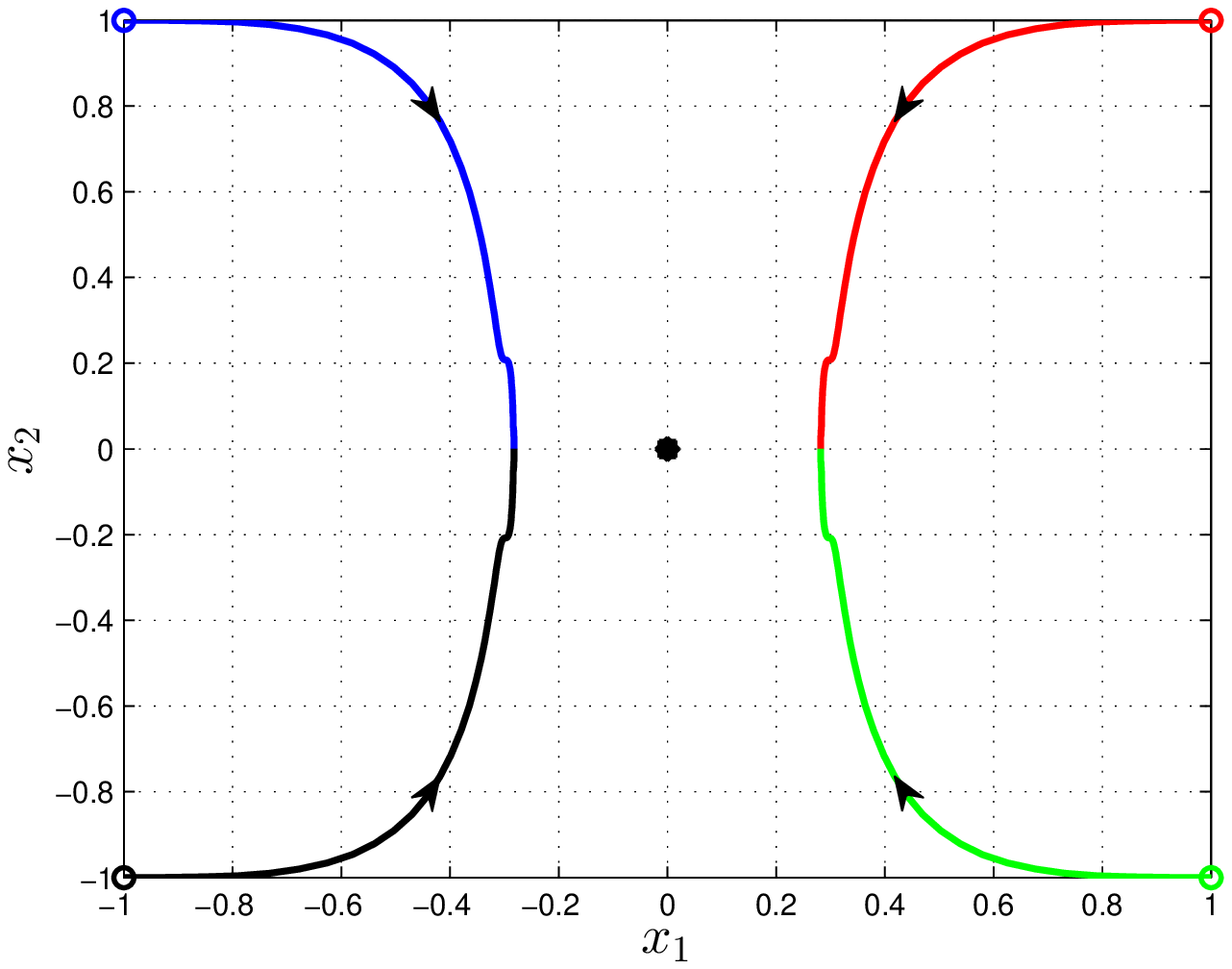}} \textit{b}
\end{minipage}
\caption{The phase trajectories of \eqref{eq7ex} for $u=0$, $d=0$ (\textit{a}) and for $u=0$, $d=1$ (\textit{b}).}
\label{Fig07aa}
\end{figure}

\textit{Example 5.}
Consider the system

\begin{equation}
\label{eq7ex}
\begin{array}{l} 
\dot{x}_1=dx_2 - x_1 x_2^2,
\\
\dot{x}_2=u-g(t)x_2,
\end{array}
\end{equation}
where $d$ takes the values of $0$ or $1$ and $g(t) = \sin^2(t)$.
It is required to design the control law $u$ that ensures the asymptotic stability of \eqref{eq7ex}. System \eqref{eq7ex} is not asymptotically stable for $u=0$ and for any values of $d$ (see Fig.~\ref{Fig07aa}).

Choose $S(x,t)=|x|^{2\alpha}$, $\alpha$ is a positive integer and use the third case of Theorem \ref{Th6}. Compute 
\begin{equation*}
\begin{array}{l}
2 \frac{\partial S(x,t)}{\partial t}+\nabla \cdot\{S(x,t)(\xi(x,t)+g(x,t)u(x,t))\}=
\\
2\alpha |x|^{2\alpha-2}(dx_1x_2-x_1^2x_2^2+ux_2-gx_2^4)+|x|^{2\alpha}(-x_2^2+\frac{\partial u}{\partial x_2}-3gx_2^2)
\end{array}
\end{equation*}
 and 
 \begin{equation*}
\begin{array}{l}
\nabla \cdot\{\xi(x,t)+g(x,t)u(x,t)\}=
\\
-2\alpha |x|^{2\alpha-2}(dx_1x_2-x_1^2x_2^2+ux_2-gx_2^4)+|x|^{-2\alpha}(-x_2^2+\frac{\partial u}{\partial x_2}-3gx_2^2).
\end{array}
\end{equation*}

\begin{figure}[h]
\begin{minipage}[h]{0.5\linewidth}
\center{\includegraphics[width=1\linewidth]{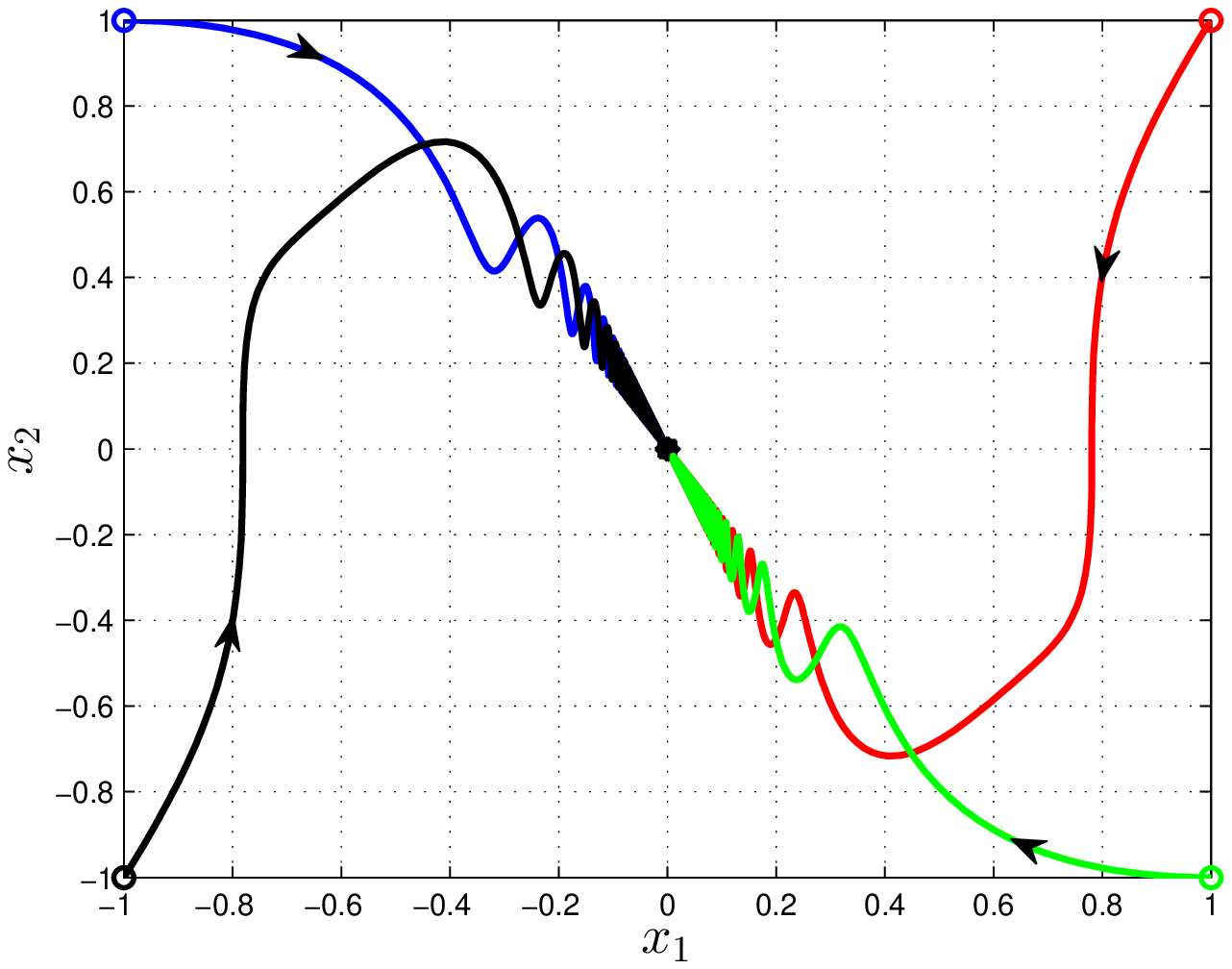}} \textit{a}
\end{minipage}
\hfill
\begin{minipage}[h]{0.5\linewidth}
\center{\includegraphics[width=1\linewidth]{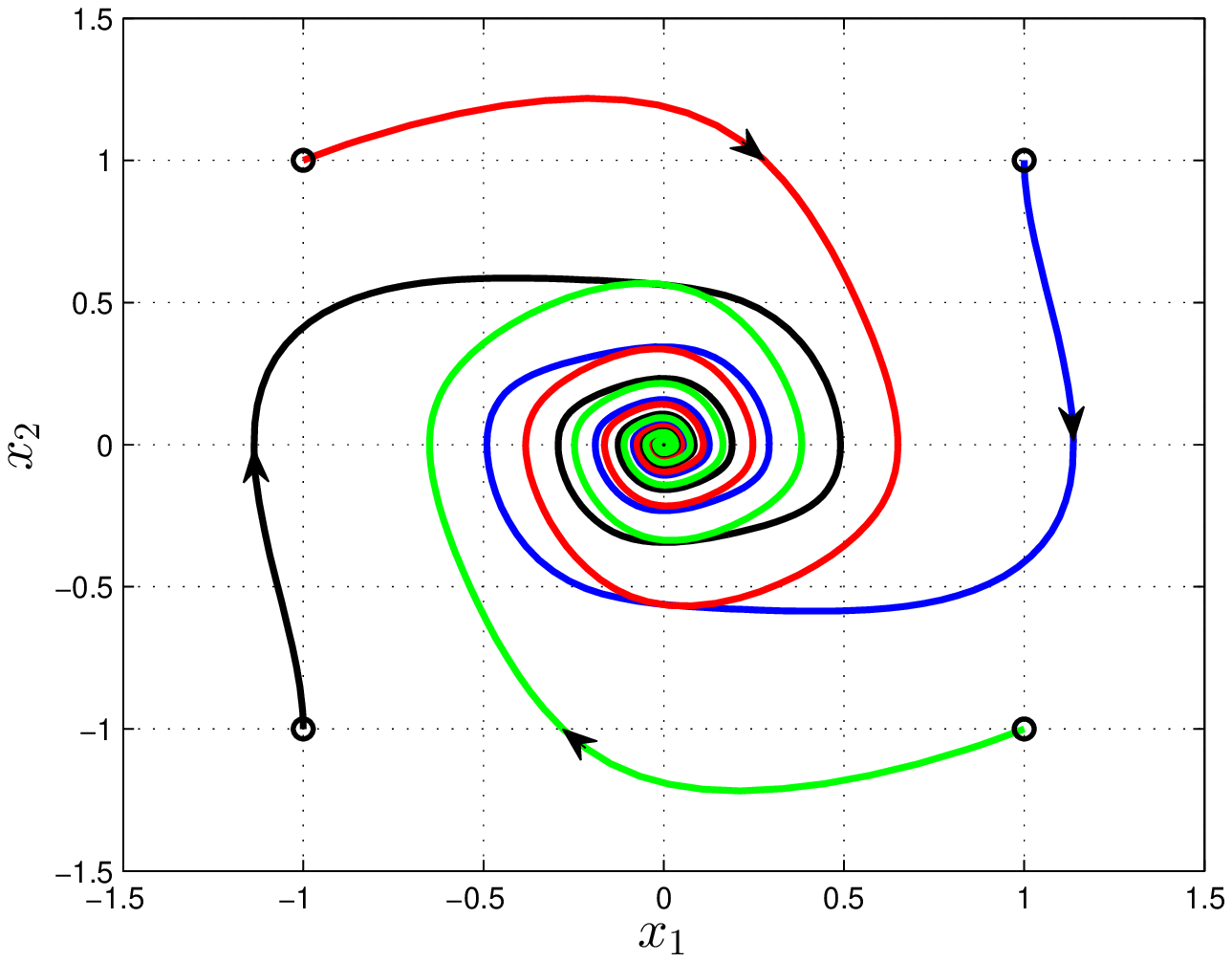}} \textit{b}
\end{minipage}
\caption{The phase trajectories in the closed-loop system for $d=0$ (\textit{a}) and for $d=1$ (\textit{b}).}
\label{Fig7aa}
\end{figure}

1. Let $d=0$. Choosing $u=-x_2^3$, we get 
$2 \frac{\partial S(x,t)}{\partial t}+\nabla \cdot\{S(x,t)(\xi(x,t)+g(x,t)u(x,t))\}<0$ and $\nabla \cdot\{\xi(x,t)+g(x,t)u(x,t)\} \leq 0$ for $x_2 \neq 0$ and any $t \geq 0$. 
The phase trajectories of the closed-loop system are shown in Fig.~\ref{Fig7aa},\textit{a}.

2. Let $d=1$. Choosing $u= -x_1-x_2^3$, we get $2 \frac{\partial S(x,t)}{\partial t}+\nabla \cdot\{S(x,t)(\xi(x,t)+g(x,t)u(x,t))\}<0$ and $\nabla \cdot\{\xi(x,t)+g(x,t)u(x,t)\} \leq 0$ for $x_2 \neq 0$ and any $t \geq 0$.
The phase trajectories of the closed-loop system are shown in Fig.~\ref{Fig7aa}, \textit{b}.

\section{Conclusion}
\label{Sec4}

A method for stability study of nonautonomous dynamical systems using the properties of the flow and divergence of the vector field is proposed. To study the stability, it is required the existence of a certain type of integration surface or the existence of an auxiliary scalar function. Necessary and sufficient stability conditions are proposed.

The obtained results are applied to synthesis the static feedback control law for dynamical systems. It is shown that the control law is found as a solution of a differential inequality, while the control law based on the method of Lyapunov functions is found as a solution of an algebraic inequality.

\section{Acknowledgments}

The results of Section 3 were developed under support of RSF (grant 18-79-10104) in IPME RAS.

\end{document}